\documentclass[12pt]{iopart}

\usepackage{graphicx}
\usepackage{bm}
\usepackage{amssymb}
\usepackage{amstext}
\usepackage{color}
\usepackage{subfigure}
\usepackage{multirow}
\begin{document}

\title[]{Experimental preparation of topologically ordered states via adiabatic evolution}

\author{Zhihuang Luo$^{1,2}$, Jun Li$^{1,2}$, Zhaokai Li$^{2}$, Ling-Yan Hung$^{3,4,5\ast}$, Yidun Wan$^{3,4,5,6\dag}$, Xinhua Peng$^{2,7\ddag}$ and Jiangfeng Du$^{2,7}$ }

\address{$^{1}$ Beijing Computational Science Research Center, Beijing, 100094, China}
\address{$^{2}$ CAS Key Laboratory of Microscale Magnetic Resonance and Department of Modern Physics, University of Science and Technology of China, Hefei, Anhui 230026, China}
\address{$^{3}$ State Key Laboratory of Surface Physics and Department of Physics, Fudan University, 220 Handan Road, 200433 Shanghai, China}
\address{$^{4}$ Department of Physics and Center for Field Theory and Particle Physics, Fudan University,
220 Handan Road, 200433 Shanghai, China}
\address{$^{5}$ Collaborative Innovation Center of Advanced Microstructures, Nanjing University,
Nanjing, 210093, China}
\address{$^{6}$ Perimeter Institute for Theoretical Physics, Waterloo, Ontario N2L 2Y5, Canada}
\address{$^{7}$ Synergetic Innovation Center of Quantum Information and Quantum Physics,
University of Science and Technology of China, Hefei, Anhui 230026, China}
\eads{\mailto{Lyhung@fudan.edu.cn}, \mailto{ydwan@fudan.edu.cn}, \mailto{xhpeng@ustc.edu.cn}}


\begin{abstract}
Topological orders are a class of exotic states of matter characterized by patterns of long-range entanglement. Certain topologically ordered systems are proposed as potential realization of fault-tolerant quantum computation. Topological orders can arise in two-dimensional spin-lattice models. In this paper, we engineer a time-dependent Hamiltonian to prepare a topologically ordered state through adiabatic evolution. The other sectors in the degenerate ground-state space of the model are obtained by applying nontrivial operations corresponding to closed string operators. Each sector is highly entangled, as shown from the completely reconstructed density matrices. This paves the way towards exploring the properties of topological orders and the application of topological orders in topological quantum memory.
\end{abstract}

%



%

\section{Introduction}

The emergence of topological order in fractional quantum Hall effect has broadened our understanding of phases of matter \cite{Wen1990TO,Wenbook,Tsui1982}. 
Phases of matter are much richer than those describable by symmetry-breaking. In the framework of Landau theory \cite{QPT,Landau1937,Ginzburg1950}, all phases of matter were thought to be characterized by different symmetries and their phase transitions were associated with broken symmetries. The discovery of distinct fractional quantum Hall states, which have exactly the same symmetries \cite{Tsui1982, Laughlin1983} were the first solid evidence that the Landau symmetry-breaking paradigm is inadequate. In particular, these phases cannot be distinguished by local order parameters. Microscopically, topological order corresponds to patterns of long-range entanglement \cite{Chen2010}. Topological entanglement entropy for example can be used as a nonlocal order parameter to (partially) identify the topological phases \cite{TEE,HammaPRB2008}.

Topological orders are of basic scientific interest not only for their topological properties in condensed matter physics \cite{Wen1990D,Arovas1984,Wen1995,TEE,TEE2} but also for their potential application in fault-tolerant quantum computation \cite{Kitaev2003,Nayak2008,Stern2013,Freedman2003}. This mainly relies on the robustness of degenerate ground states \cite{Wen1990D} and their patterns of long-range entanglement \cite{Chen2010}. Topologically ordered states support degenerate ground states when the system is placed in a geometry with non-trivial topology. In two dimensions,  its ground-state degeneracy depends on the genus of the Riemann surface it lives in. Because any two orthogonal sectors in ground-state sub-space are related by non-contractible loop operators, local perturbations only lead to local errors that can be easily detected and corrected. This translates into effective correction of both phase flip and bit flip errors in explicit models. These properties are attractive for constructing a robust memory register \cite{Dennis2002,Jiang2008}.

Two-dimensional spin-lattice models exhibit topological orders \cite{Kitaev2003,Kitaev2006,WenPRL2003}. However, such models usually include many-body interactions that have not been found in natural physical systems. Feynman suggested that \cite{Feynman1982} a well-controlled quantum system can be used for the efficient simulation of other quantum systems. This  implies that one can bypass limitations set by natural materials, and study detailed quantum phenomenon of exotic systems via quantum simulations. Quantum simulation could be realized by various physical platforms such as trapped ion \cite{Friedenauer2008}, superconducting circuit \cite{Clarke2008}, nuclear magnetic resonance (NMR) \cite{Chuang2005}, and so on. There have been a lot of successful demonstrations in simulating condensed matter physics \cite{Peng2005, Kim2010, Kandala2017}, high-energy physics \cite{Li2017measuring}, atomic physics \cite{Bernien2017}, quantum chemistry \cite{Du2010, Lanyon2010}, and cosmology \cite{Nation2009, QSRMP}. Quantum simulation thus provides a powerful tool to investigate topological order \cite{Luo2014, Luo2016, Luo2018, Kong2016, Li2017}.

In this paper, we realize the Hamiltonian of Wen-plaquette spin-lattice model with $Z_2$ topological orders in an NMR system. There are four-fold degenerate ground states when the Hamiltonian is chosen to have doubly periodic boundary condition, i.e., we have placed the system on a torus. Following the method proposed by Hamma \textit{et al}. \cite{HammaPRL2008}, a topologically ordered state is experimentally prepared through adiabatically engineering the time-dependent Hamiltonian. The other sectors in the degenerate ground-state subspace are obtained by performing two nontrivial closed string operations. We reconstruct the density matrices for each sector using complete quantum state tomography technology. The results show that each sector exhibits maximum entanglement. It is a central step to prepare degenerate topologically ordered states for studying topological quantum phase transition \cite{Luo2014}, measuring modular matrices \cite{Luo2018}, constructing robust quantum memory, and so on.

\section{Wen-plaquette model}

\begin{figure}
  \centering
  \includegraphics[width=0.6\linewidth]{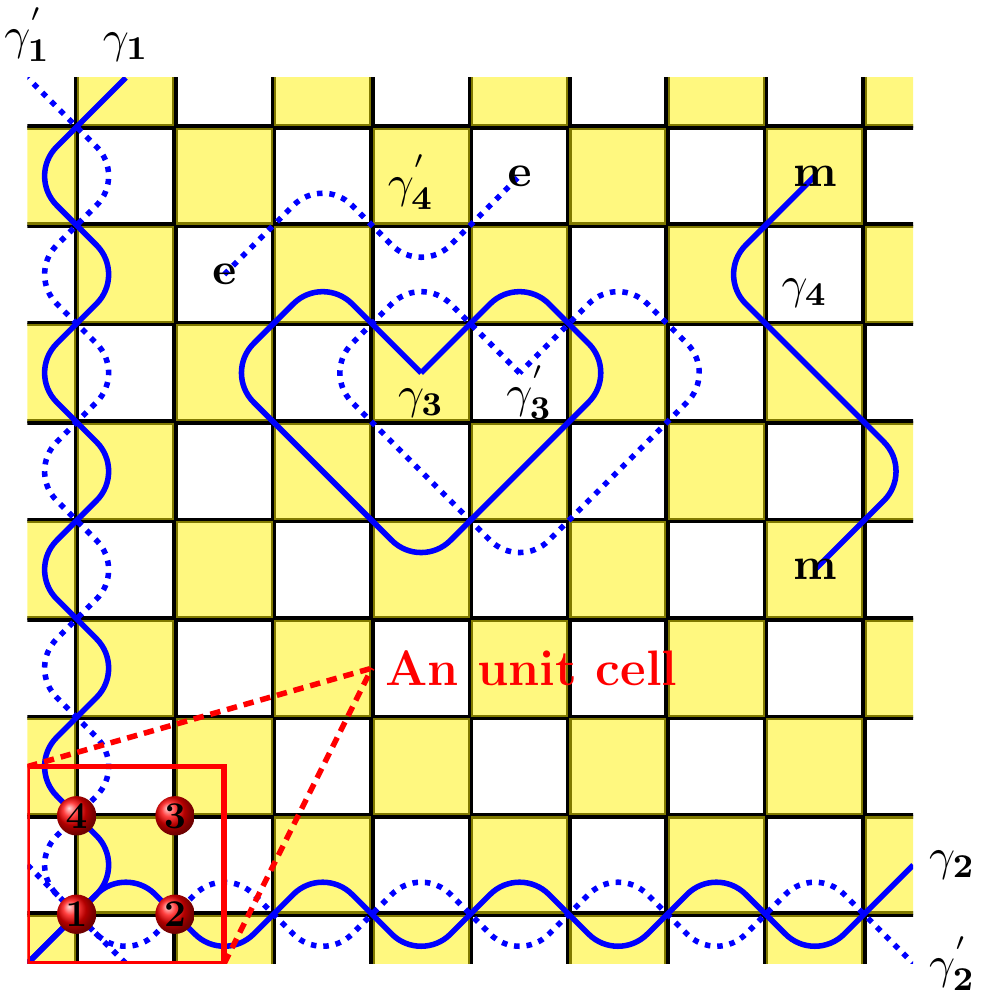}
  \caption{The Wen-plaquette spin-lattice model on a torus. When $N$ is even, there exist two sublattice denoted by white and yellow. The blue solid strings ($\gamma_1\sim\gamma_4$) and their dual dashed strings ($\gamma_1^{'}\sim\gamma_4^{'}$) are defined in the yellow sublattice and in the white sublattice, respectively. $e$ and $m$ represent the elementary excitations (anyons): electric charge and magnetic vortex, which in pairs generated by open string operators. The red box is an unit cell of square lattice. The red balls represent spins placing on each site.}\label{fig:model_1}
\end{figure}

The Wen-plaquette model building on an $N\times N$ square lattice is illustrated in figure \ref{fig:model_1}, where each site accommodates a spin $1/2$ \cite{WenPRL2003}. Its Hamiltonian is written as
\begin{equation}\label{Ham}
    H_{\textrm{Wen}}=-\sum_{\textrm{white plaquettes}}X_p-\sum_{\textrm{yellow plaquettes}}Z_{p}.
\end{equation}
Here $X_p=\prod_{j\in\partial p}\sigma_j^x, Z_p=\prod_{j\in\partial p}\sigma_j^z$ are the plaquette operators that act on four spins surrounding a plaquette $p$, and $\sigma_j^{\alpha}$'s stand for Pauli operators. This is an exactly solvable model because $[X_p,Z_{p'}]=0$ for all $p$ and $p'$. One can show that both the Wen-plaquette model and the toric code model describe the $Z_2$ topological order. The ground state manifold $\mathcal{L}$ is given by
\begin{equation}\label{}
  \{|\psi_g\rangle\in\mathcal{H}:X_p|\psi_g\rangle=Z_p|\psi_g\rangle=|\psi_g\rangle \textrm{ for all } p\}.
\end{equation}
The  ground state degeneracy $D$ depends on the genus $\textsf{g}$ of the Riemann surface, i.e., $D=2^{2\textsf{g}}$. If the Riemann surface has genus $\textsf{g}$, we can define $2\textsf{g}$ non-contractible strings that connect different topological sectors in $\mathcal{L}$. For example, on a torus its four-fold degenerate ground states can be described by using two nontrivial closed strings of $\gamma_1$ and $\gamma_2$,
\begin{equation}\label{GS}
    |\psi_g^{(\nu_1,\nu_2)}\rangle=\mathcal{T}_x^{\nu_1}(\gamma_1)\mathcal{T}_x^{\nu_2}(\gamma_2)|\psi_g^{(0,0)}\rangle, \nu_1,\nu_2=0,1.
\end{equation}
Here the string operators are defined as $\mathcal{T}_x(\gamma)=\prod_{j\in \gamma}\sigma_j^x$. The initial topological sector $|\psi_g^{(0,0)}\rangle$ of $\mathcal{L}$ is given by the equal superposition of all contractible closed strings including no string (or say the string is a point). It is not difficult to find that $\mathcal{T}_x(\gamma_3)=\prod_{p\in f}X_p$, namely, the product of all plaquette operators in the surface $f$ that satisfies $\partial f=\gamma_3$. And for the case of no string, its corresponding operator is the identity matrix $I$. All operators corresponding to such strings form a group denoted as $\mathcal{G}$. Its elements are generated by $n=(N^2-2)/2$ independent plaquette operators in the white sublattice, i.e., $g_s=\prod_{p=1}^nX_p^{s_p}, s_p\in\{0,1\}$. Thus,
\begin{equation}\label{GS0}
|\psi_g^{(0,0)}\rangle=\frac{1}{\sqrt{|\mathcal{G}|}}\sum_{g_s\in\mathcal{G}}^{|\mathcal{G}|}g_s|00\cdots\rangle,
\end{equation}
where $|\mathcal{G}|$ is the number of elements in $\mathcal{G}$. Also, $|0\rangle$ stands for spin-up states along the $z$ axis. According to the closure of group and the commutation relations, we can prove that $X_p|\psi_g^{(\nu_1,\nu_2)}\rangle=Z_p|\psi_g^{(\nu_1,\nu_2)}\rangle=|\psi_g^{(\nu_1,\nu_2)}\rangle$ for all $p,\nu_1,\nu_2$. Therefore, the states of $|\psi_g^{(\nu_1,\nu_2)}\rangle$ constructed from equations (\ref{GS}) and (\ref{GS0}) are indeed the ground states of the Hamiltonian (\ref{Ham}). Besides, it shows that $\langle\psi_g^{(\nu_1,\nu_2)}|\psi_g^{(\nu_1^{'},\nu_2^{'})}\rangle=\delta_{\nu_1,\nu_1^{'}}\delta_{\nu_2,\nu_2^{'}}$, meaning that, different topological sectors are orthogonal.

A unit cell on a torus is illustrated in the red box of figure \ref{fig:model_1}, which consists of $2\times 2$ spins. Although it is the smallest system, the properties of topological orders are still presented in the Wen-plaquette spin-lattice model since they describe fixed wavefunction \cite{WenPRL2003}. This validity also follows from the fairly short-range spin-spin correlations \cite{Luo2014}. Under the periodic boundary condition, we can get the Hamiltonian
\begin{equation}
\hat{H}_{\textrm{Wen}}^{4}=-2(X_p+Z_p).
\end{equation}
Two nontrivial loop operators are $\mathcal{T}_x(\gamma_1)=\sigma_1^x\sigma_4^x$ and $\mathcal{T}_x(\gamma_2)=\sigma_1^x\sigma_2^x$ and all contractible closed-string operators form a group, i.e., $\mathcal{G}=\{I,X_p\}$, which are generated by one independent plaquette operator, i.e., $X_p$. So from equations (\ref{GS}) and (\ref{GS0}), the four-fold degenerate ground states can be described as follows,
\begin{eqnarray}\label{GS1}
    |\psi_g^{(0,0)}\rangle =(|0000\rangle+|1111\rangle)/\sqrt{2} ~\\ \nonumber
    |\psi_g^{(0,1)}\rangle =(|0011\rangle+|1100\rangle)/\sqrt{2} ~\\ \nonumber
    |\psi_g^{(1,0)}\rangle =(|0110\rangle+|1001\rangle)/\sqrt{2} ~\\ \nonumber
    |\psi_g^{(1,1)}\rangle =(|0101\rangle+|1010\rangle)/\sqrt{2}. 
\end{eqnarray}
Here we work with basis states which are eigenstates of $\sigma_z$ at each site. The theory shows that each topological sector in equation (\ref{GS1}) is a maximally entangled state through stochastic local quantum operation assisted by classical communication (SLOCC) \cite{Verstraete2002, Regula2014}.

\section{Experiment}

\begin{figure}
  \centering
  \includegraphics[width=0.8\linewidth]{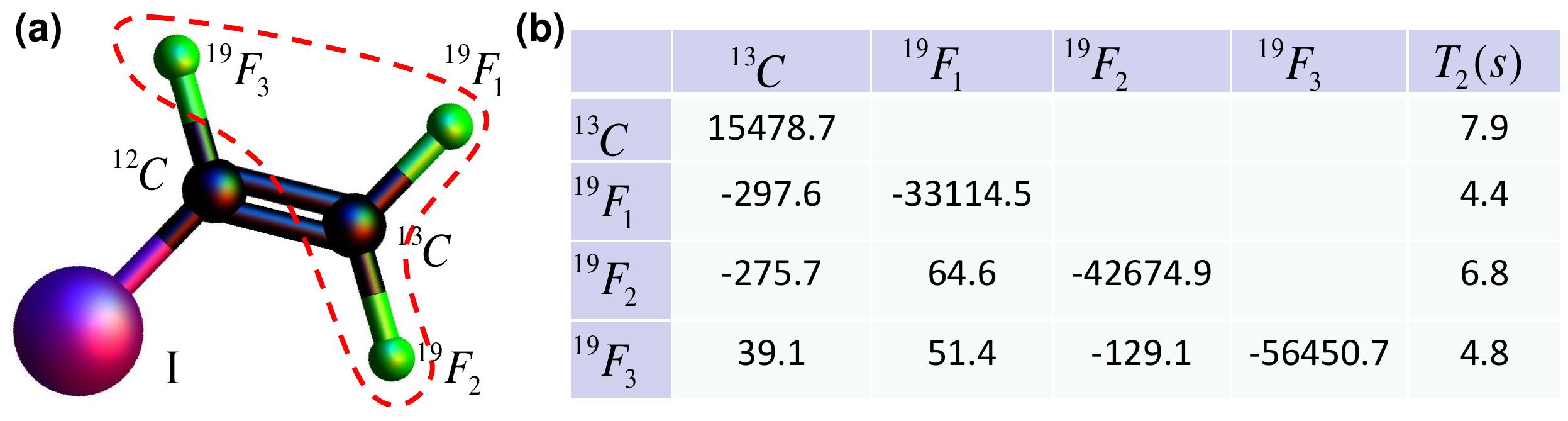}
  \caption{ (a) Molecular structure of lodotrifluroehtylene, where $^{13}$C and three $^{19}$F nuclei are used as a 4-qubit quantum simulator. (b) The chemical shifts and the coupling constants (in units of Hz) are on and below the diagonal in the table.The spin-lattice relaxation times ($T_1$) are  $21$ s for $^{13}$C and $12.5$ s for $^{19}$F.}\label{fig:sample}
\end{figure}

The experiment was carried out on a Bruker AV-400 spectrometer ($9.4T$) at room temperature $T=300$ K. We chose one $^{13}$C  and three $^{19}$F nuclear spins of Iodotrifluroethylene (C$_2$F$_3$I) as a 4-qubit quantum simulator. Its molecular structure and relevant parameters are shown in figures \ref{fig:sample}(a) and \ref{fig:sample}(b). The natural Hamiltonian of this system under weak coupling approximation is
\begin{equation}
\hat{H}_{\textrm{NMR}}=\sum_{j=1}^{4}\frac{\omega _{j}}{2}\hat{\sigma}
_{z}^{j}+\sum_{j<k,=1}^{4}\frac{\pi J_{jk}}{2}\hat{\sigma} _{z}^{j}\hat{%
\sigma} _{z}^{k},
\end{equation}%
where $\omega _{j}$ represents the chemical shift of spin $j$ and $J_{jk}$ the coupling
constant between spin $j$ and spin $k$.  

\begin{figure}
  \centering
  \includegraphics[width=0.8\linewidth]{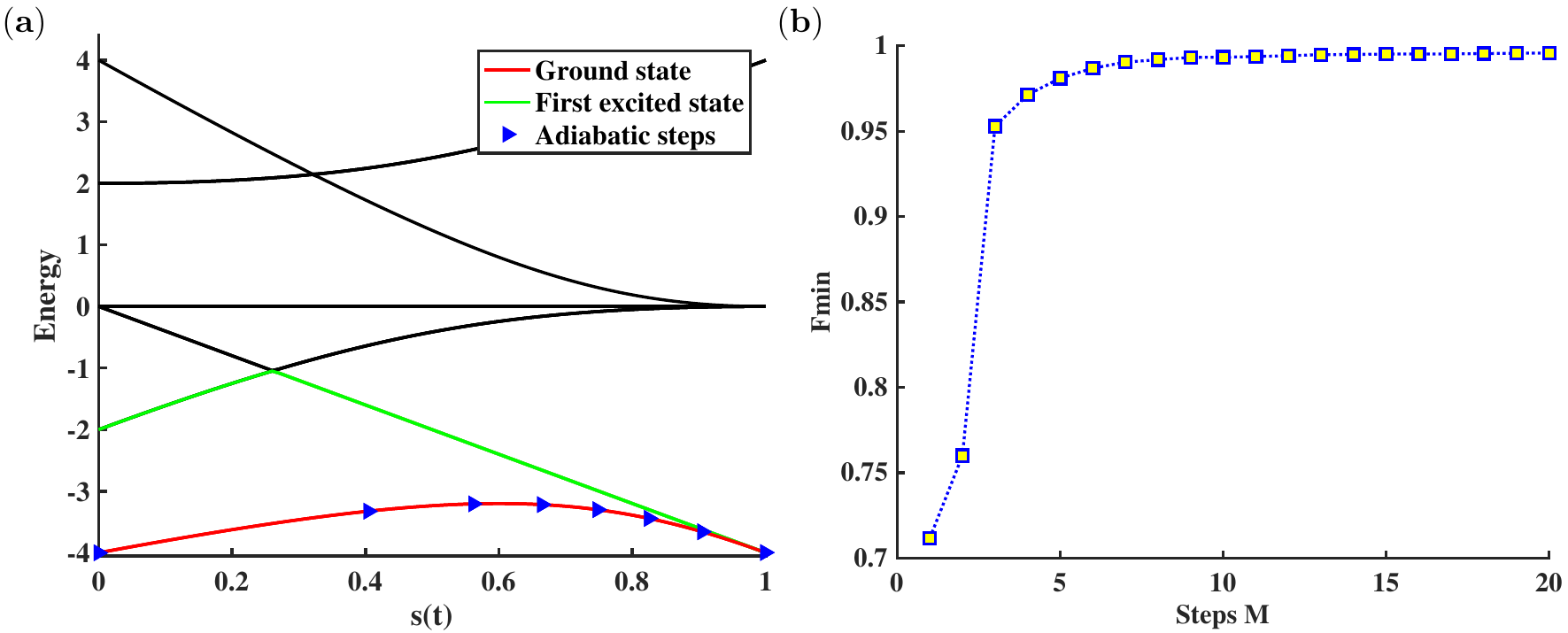}
  \caption{(a) The energy-level diagram of the time-dependent Hamiltonian $\hat{H}[s(t)]$. The red and green curves represent the energies of ground state and first excited state, respectively. The blue $\vartriangleright$ points represent the $M=7$ interpolations on the line of $s(t)$ with the same time interval $\tau$. (b) The minimum fidelity $F_{\textrm{min}}$ during the adiabatic passage as the function of the number of steps $M$.}\label{fig:energy}
\end{figure}

The quantum system was firstly prepared in the initial pseudo-pure state (PPS):$\hat{\rho}_{0000}=\frac{1-\epsilon}{16}\bold{I}+\epsilon |0000\rangle\langle0000|$ using line-selective approach \cite{Peng2001}, with $\bold{I}$ representing $16\times16$ identity operator and $\epsilon\approx 10^{-5}$ the polarization. Note that $|0000\rangle$ is  the ground state of $\hat{H}_0=-\sum_{j=1}^4\hat{\sigma}_j^z$. We then adiabatically prepare the ground state $|\psi_g^{(0,0)}\rangle$ of Wen-plaquette model by varying the time-dependent Hamiltonian sufficiently slowly,
\begin{equation}\label{Had}
    \hat{H}(t)=[1-s(t)]\hat{H}_0+s(t)\hat{H}_{\textrm{Wen}}^4,
\end{equation}
where the parameter function $s(t)$ increases monotonically from 0 at $t=0$ to 1 at $t=T$. The energy levels of $\hat{H}(t)$ as the function of the parameter $s(t)$ are shown in figure \ref{fig:energy}(a). The red and green curves represent the ground-state and first excited-state energies, respectively. To ensure that the system is prepared to the  ground state of target Hamiltonian at $t=T$, the variation of the control parameter has to be slow sufficiently, i.e., satisfing the adiabatic condition \cite{Messiah1976},
\begin{center}
\begin{equation}
   \frac{\langle \psi _{g}|\frac{\partial \hat{H}(t)}{\partial s(t)}\frac{\partial s(t)}{\partial t}|\psi _{e}\rangle }{(\varepsilon_{e}-\varepsilon _{g})^{2}}=\epsilon\ll 1.
\end{equation}
\end{center}
This condition determines the optimal sweep of control parameter $s(t)$, which was interpolated with M discretized scan steps.  The duration of each step is defined as $\tau=T/M$. Thus the adiabatic condition is satisfied when both $T, M \to \infty$ and  $\tau \to 0$. To determine the optimal $M$  in the adiabatic transfer, we numerically simulated the minimum fidelity $F_{\textrm{min}}$ encountered during the scan as a function of the number of steps, as shown in figure \ref {fig:energy}(b), where we fixed the total evolution time $T=2.9982$. The fidelity is calculated as the overlap of the theoretical and simulated ground states at each step of adiabatic evolution. In the experiment, we discretized into $M=7$ steps and the minimal fidelity $F_{\textrm{min}}$ is $\geq 0.99$, which indicates that the state of the system is always close to its instantaneous ground state in the whole adiabatic passage.

\begin{figure}
\centering
\includegraphics[width=0.8\linewidth]{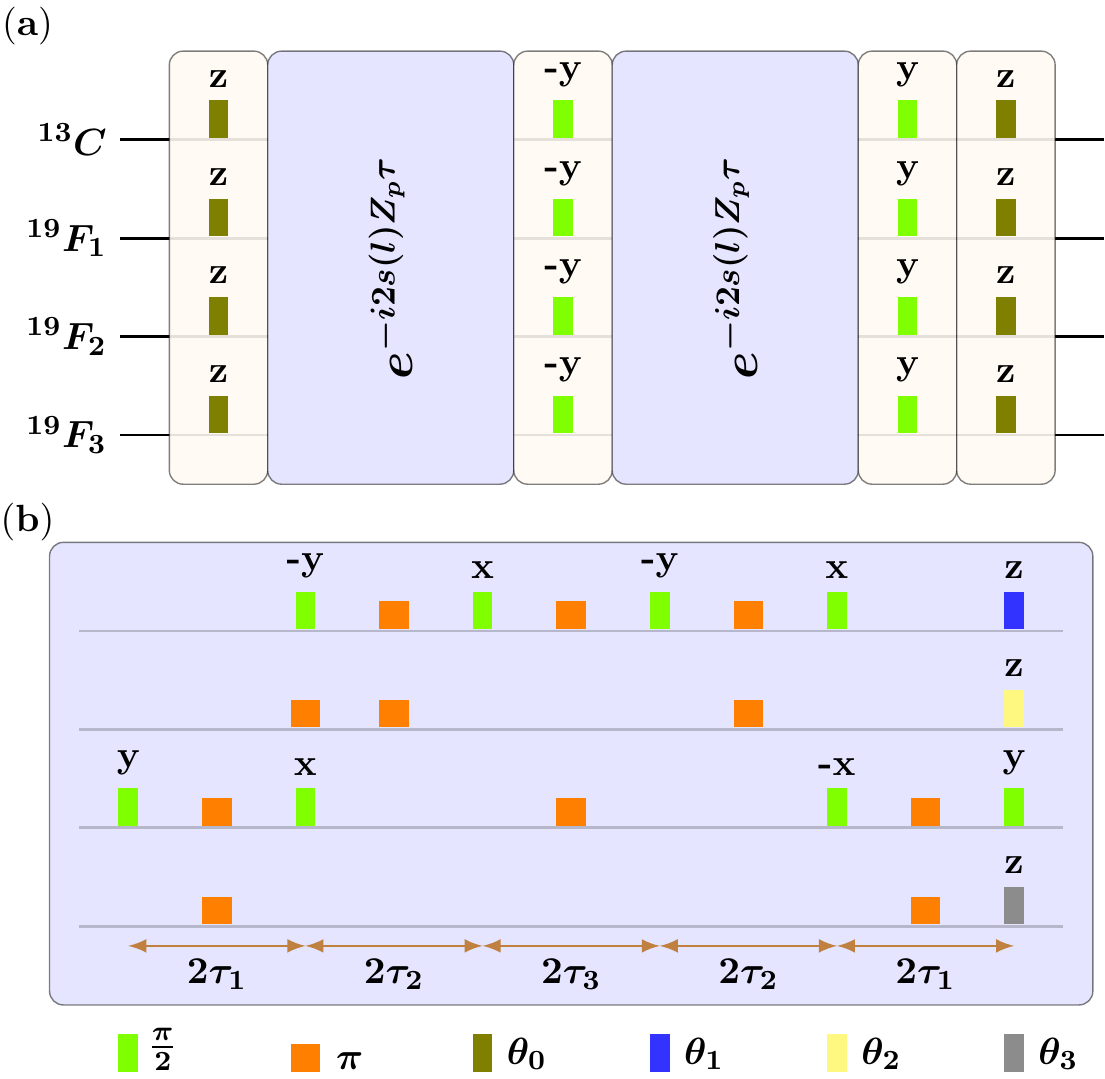}
\caption{Pulse sequences for (a) implementing the adiabatic evolution of $l_{th}$ step, and (b) effectively creating four-body interaction, i.e., $Z_p=\hat{\protect \sigma}_z^1\hat{\protect \sigma}_z^2\hat{\protect \sigma}_z^3\hat{\protect \sigma}_z^4$, where $\tau_1=1/4J_{34},\tau_2=1/4J_{12},\tau_3=2s(l)\tau/\pi J_{13},\theta_0=[1-s(l)]\tau/2,\theta_1=-\omega_1/J_{34},\theta_2=-4\omega_2s(l)\tau/\pi J_{13}$ and $\theta_3=\omega_4/J_{12}+4\omega_4s(l)\tau/\pi J_{13}$.}
\label{fig:pulseq}
\end{figure}

The adiabatic evolution for each step can be decomposed by Suzuki-Trotter expansion as
\begin{equation}\label{eq:ps1}
    e^{-i\hat{H}[s(l)]\tau} =e^{-i[1-s(l)]\hat{H}_0\tau/2}e^{-i s(l)\hat{H}_{\textrm{Wen}}^4\tau} e^{-i[1-s(l)]\hat{H}_0\tau/2}+ O(\tau^3),
\end{equation}
with $l=0,1,\cdots,M$. This expansion faithfully represents the targeted evolution if the duration $\tau$ provided is kept sufficiently short. Due to $[X_p,Z_p] =0 $, we have
\begin{equation}\label{eq:ps2}
    e^{-is(l)\hat{H}_{\textrm{Wen}}^4\tau}=\prod_{j=1}^4\hat{R}_j^y(\frac{\pi}{2}) e^{i 2 s(l) Z_P \tau}\prod_{j=1}^4\hat{R}_j^y(-\frac{\pi}{2}) e^{ i 2 s(l) Z_p \tau}.
\end{equation}
Here the four-body interaction $Z_p$ is effectively created by a combination of RF pulses and free evolutions of NMR system \cite{Tseng1999, Peng2009}:
\begin{eqnarray}\label{eq:ps3}
\fl
e^{-i 2 s(l) Z_p \tau }
=\hat{R}_1^z(\theta_1)\hat{R}_2^z(\theta_2)\hat{R}_4^z(\theta_3)\hat{R}_3^y(\frac{\pi}{2})e^{-i\hat{H}_{\textrm{NMR}}\tau_1} \hat{R}_3^y(\pi)\hat{R}_4^y(\pi)e^{-i\hat{H}_{\textrm{NMR}}\tau_1}\hat{R}_1^y(-\frac{\pi}{2})\\ \nonumber
\cdot \hat{R}_3^x(-\frac{\pi}{2}) e^{-i\hat{H}_{\textrm{NMR}}\tau_2}\hat{R}_1^y(\pi)\hat{R}_2^y(\pi)e^{-i\hat{H}_{\textrm{NMR}}\tau_2}\hat{R}_1^x(\frac{\pi}{2})e^{-i\hat{H}_{\textrm{NMR}}\tau_3} \hat{R}_1^x(\pi)\hat{R}_3^x(\pi)\\ \nonumber
\cdot e^{-i\hat{H}_{\textrm{NMR}}\tau_3}\hat{R}_1^x(\frac{\pi}{2})e^{-i\hat{H}_{\textrm{NMR}}\tau_2}\hat{R}_1^y(\pi) \hat{R}_2^y(\pi)e^{-i\hat{H}_{\textrm{NMR}}\tau_2}\hat{R}_1^y(-\frac{\pi}{2})\hat{R}_3^x(\pi/2) \\ \nonumber
\cdot \hat{R}_2^y(\pi)e^{-i\hat{H}_{\textrm{NMR}}\tau_1} \hat{R}_3^x(\pi)\hat{R}_4^x(\pi) e^{-i\hat{H}_{\textrm{NMR}}\tau_1}\hat{R}_3^y(\frac{\pi}{2}),
\end{eqnarray}
where $\hat{R}_j^{\alpha}(\theta)=e^{-i\theta\hat{\sigma}_j^{\alpha}/2}(\alpha=x,y,z)$. Therefore, the adiabatic evolution for each step can be implemented by the pulse sequences, as shown in figures \ref{fig:pulseq}(a) and \ref{fig:pulseq}(b). This simulation method is in principle efficient, provided the decoherence time is long enough. In order to overcome the accumulated pulse errors and the decoherence, each step of adiabatic evolution $e^{-i\hat{H}[s(l)]\tau}$ ($l= 0, 1, 2, ...,M $) was optimized by the gradient ascent pulse engineering (GRAPE) algorithm \cite{Glaser2005}. The GRAPE pulses of $e^{-i\hat{H}[s(l)]\tau}$ were designed to have the pulse length of 30 $ms$ and theoretical fidelity of over 0.99 in experiments.

\begin{figure}
  \centering
  \includegraphics[width=0.8\linewidth]{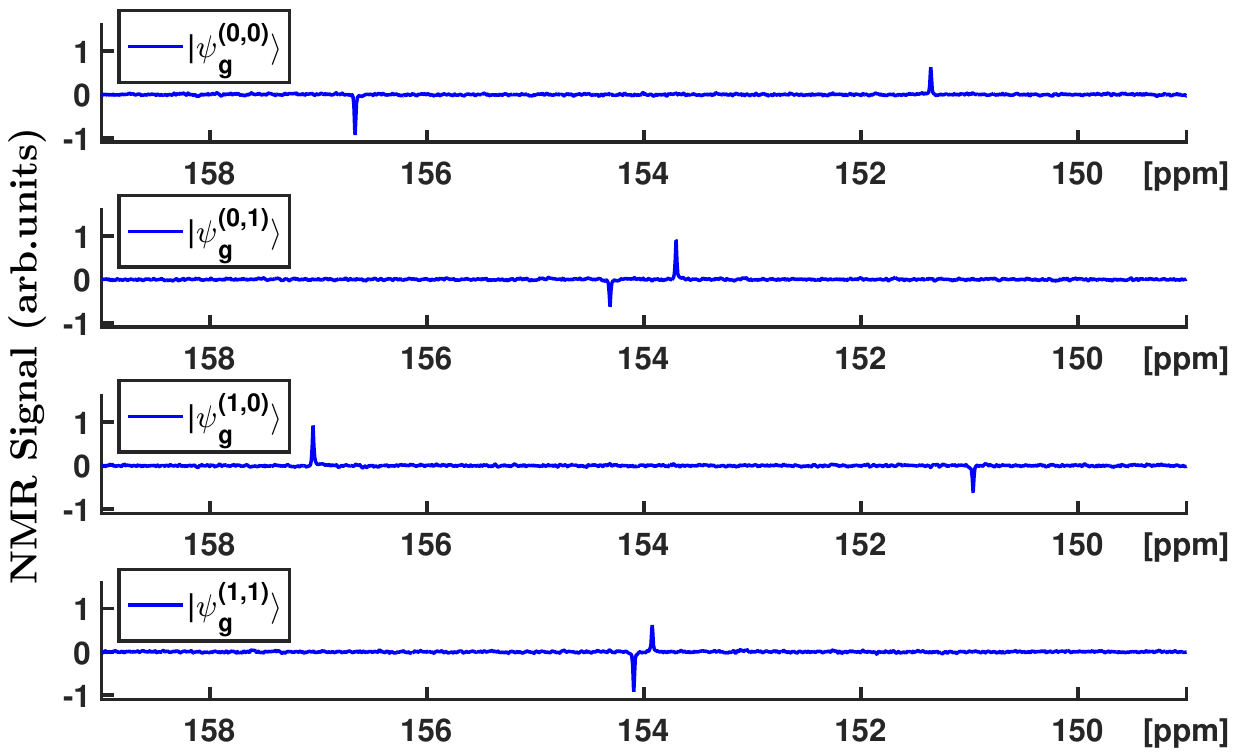}\\
  \caption{Experimental $^{13}C$ spectra corresponding to four topological sectors i.e., $|\psi_g^{(0,0)}\rangle,|\psi_g^{(0,1)}\rangle,|\psi_g^{(1,0)}\rangle$, and $|\psi_g^{(1,1)}\rangle$ from top to bottom, respectively.}\label{fig:spec}
\end{figure}

\begin{figure}
  \centering
  \includegraphics[width=0.95\linewidth]{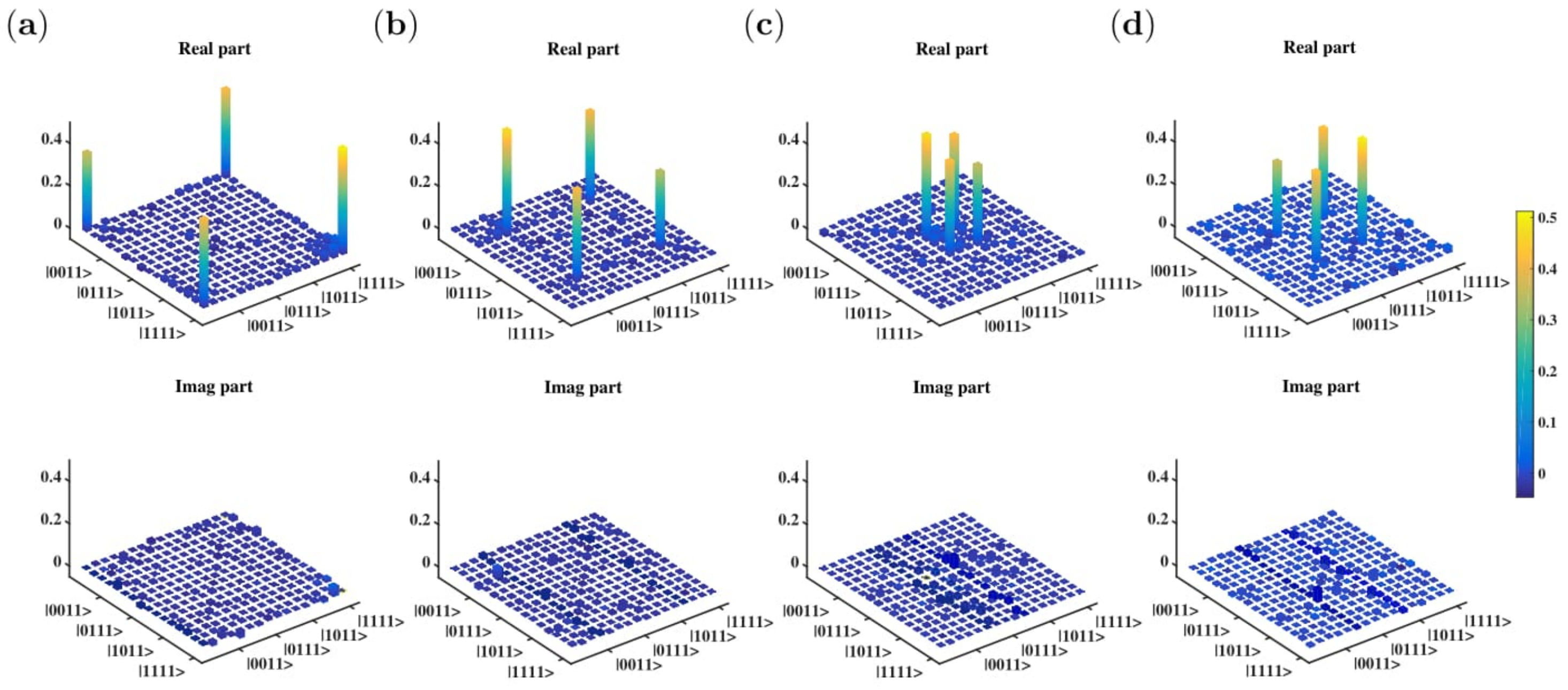}\\
  \caption{Experimental reconstructed density matrices for different topological sectors, that is, $|\psi_g^{(0,0)}\rangle,|\psi_g^{(0,1)}\rangle,|\psi_g^{(1,0)}\rangle$, and $|\psi_g^{(1,1)}\rangle$ corresponding to (a)$\sim$(d), respectively. The top an bottom represent the real parts and imaginary parts.}\label{fig:tomo2}
\end{figure}

Once the initial topologically ordered state $|\psi_g^{(0,0)}\rangle$ was prepared via the adiabatic evolution, the other topological sectors in $\mathcal{L}$ were obtained by performing non-contractible string operators $\mathcal{T}_x^{\nu_1}(\gamma_1)\mathcal{T}_x^{\nu_2}(\gamma_2)$ on $|\psi_g^{(0,0)}\rangle$. Due to $\langle\psi_g^{(\nu_1,\nu_2)}|\dot{H}|\psi_g^{(\nu_1^{'},\nu_2^{'})}\rangle=0$ ($\nu_1\neq \nu_1^{'}$ or $\nu_2\neq \nu_2^{'}$), the transition between different topological sectors are forbidden during the adiabatic evolution.  These nontrivial operators make it possible to have experimental access to different topological sectors. The resulting $^{13}C$ spectra for different quantum states are illustrated in figure \ref{fig:spec}, after a readout pulse $\hat{R}_1^y(\frac{\pi}{2})$ acting on $^{13}C$ observable nucleus. Figures \ref{fig:spec}(a) $\sim$ \ref{fig:spec}(d) correspond to the spectra of $|\psi_g^{(0,0)}\rangle,|\psi_g^{(0,1)}\rangle,|\psi_g^{(1,0)}\rangle$, and $|\psi_g^{(1,1)}\rangle$, respectively. To further confirm the experiments, we reconstructed the quantum state density matrices using the complete tomography technology \cite{Lee}. The coefficients of 256 operators for a full density matrix of four-qubit state can be obtained by performing 44 independent experiments. These readout pulses are shown in Table.~\ref{tab:tomography1}, which involve 28 local rotations and 3 SWAP gates. They were realized by GRAPE pulses with the length of 400 $\mu s$ for local rotations, 9 $ms$ for SWAP gates between $^{13}C$ and $F_1$, $F_2$, and 30 $ms$ for SWAP gate between $^{13}C $ and $F_3$, respectively. The reconstructed results are shown in figure \ref{fig:tomo2}, with the fidelities being $96.46\%,96.59\%,96.06\% \textrm{ and } 96.06\%$ for four topological sectors in $\mathcal{L}$, respectively. The receivable fidelities ensure that it is successful to adiabatically prepare the topological orders with patterns of long-range entanglement.

\begin{table}
\caption{Readout pulses for fully reconstructing a density matrix of 4-qubit state. Here E is the unit operator; X and Y denote a $\pi/2$ rotation along $x$ and $y$ axis; SWAP$_{ij}$ represents a SWAP gate between spin $i$ and $j$ in order to transfer the $^{19}$F information to $^{13}$C, and then all signals are obtained from the $^{13}$C spectra.}
\label{tab:tomography1}
\footnotesize
\centering
\begin{tabular}{c}
\hline\hline
   EEEE, EXEE,  EYEE, EEXE, EXXE, EYXE, EEYE, EXYE, EYYE, EEEX, EXEX, \\
   EYEX, EEXX,  EXXX, EYXX, EEYX, EXYX, EYYX, EEEY, EXEY, EYEY, EEXY, \\
   EXXY, EYXY, EEYY, EXYY, EYYY, YEEE, EEEE*$\textrm{SWAP}_{12}$, EEXE*$\textrm{SWAP}_{12}$, \\
   EEYE*$\textrm{SWAP}_{12}$, EEEX*$\textrm{SWAP}_{12}$, EEXX*$\textrm{SWAP}_{12}$, EEYX*$\textrm{SWAP}_{12}$, EEEY*$\textrm{SWAP}_{12}$, \\
   EEXY*$\textrm{SWAP}_{12}$, EEYY*$\textrm{SWAP}_{12}$, EEE*$\textrm{SWAP}_{12}$, EEEE*$\textrm{SWAP}_{13}$, EEEX*$\textrm{SWAP}_{13}$, \\
   EEEY*$\textrm{SWAP}_{13}$, YEEE*$\textrm{SWAP}_{13}$, EEEE*$\textrm{SWAP}_{14}$, YEEE*$\textrm{SWAP}_{14}$. \\
 \hline\hline
\end{tabular}
\end{table}

These experimental results are in good agreement with theoretical expectations. The relatively minor deviations can be attributed mostly to the imperfections of adiabatic approximation, GRAPE pulses and the spectral integrals. The theoretical infidelities of adiabatic approximation and GRAPE pulses are around $1\%$. Taking both errors into account, the numerical simulation gives the fidelities of $97.59\%,97.96\%,97.56\% \textrm{ and } 97.75\%$ for four topological sectors. Though we used the spectral fitting method in experiments, there are about $1\%\sim1.5\%$ errors from the spectral integrals by the comparison of the simulated and experimental results.

\section{Discussion and conclusion}

The Wen-plaquette spin-lattice model with $Z_2$ topological order is supported by four-body interactions, which is very rare and yet to be found in naturally occurring systems. Quantum simulation provides a powerful means to investigate these systems. Using an NMR simulator, we realized the model on a  $2\times 2$ lattice on a torus. A topologically ordered state was experimentally prepared through adiabatically engineering the time-dependent Hamiltonian and the other sectors in the degenerate ground-state manifold were obtained by performing two non-contractible string operations. The experimental results were confirmed by the complete quantum state tomography. From the reconstructed density matrices, it shows that each topological sector is maximally entangled. The experiment demonstrated the feasibility of adiabatic method to prepare topological orders. In Ref. \cite{HammaPRL2008}, it was also shown that the adiabatic timescale T scales at worst only as $\sqrt{n}$ for a system of $n$ spins, which means that we can apply the method to larger and more generic systems.  The successful preparation of different topological sectors in degenerate ground-state manifold is crucial towards further study of robust properties of topological phase, and initial construction of topologically fault tolerant quantum memory. 

\section*{Acknowledgments}

This work is supported by NKBRP(2013CB921800 and 2014CB848700), the National Science Fund for Distinguished Young Scholars (11425523), NSFC(11375167, 11227901,11734002, 11374032 and 91021005), the Strategic Priority Research Program (B) of the CAS(XDB01030400), and RFDP (20113402110044). YW acknowledges the support from the John Templeton foundation No. 39901. This research was supported in part by Perimeter Institute for Theoretical Physics. Research at Perimeter Institute is supported by the Government of Canada through the Department of Innovation, Science and Economic Development Canada and by the Province of Ontario through the Ministry of Research, Innovation and Science.  LYH would like to acknowledge support by the Thousand Young Talents Program, and Fudan University.


\end{document}